# Ghost Imaging with Blackbody Radiation

Yangjian Cai and Shi-Yao Zhu

*Department of Physics, Hong Kong Baptist University, Hong Kong, China*

*Institute of Optics, Department of Physics, ZheJiang University, Hangzhou, 310027, China*

Abstract

We present a theoretical study of ghost imaging by using blackbody radiation source. A Gaussian thin lens equation for the ghost imaging, which depends on both paths, is derived. The dependences of the visibility and quality of the image on the transverse size and temperature of the blackbody are studied. The main differences between the ghost imaging by using the blackbody radiation and by using the entangled photon pairs are image-forming equation, and the visibility and quality of the image.

Ghost imaging were firstly realized by using entangled photon pairs generated in spontaneous parametric down conversion in 1995 [1,2]. The name ghost comes from the facts that an object in one path produces an image in another path in the measurement of coincident counting rates, and the image depends on both paths. Since then, many theoretical and experimental studies on this subject have been published [3-10]. Recently, there were discussions about whether quantum entanglement is necessary in the ghost imaging and whether a ghost imaging experiment can be realized with classical source [11-14]. Bennink et al. presented their classical ghost imaging and interference experiments [12-13]. However, they did not give the imaging formation equation. Furthermore, Angelo et al. [14] claimed that the classical ghost image [12] is a shot by shot, point-to-point projection. Gatti et al first pointed out theoretically that the ghost imaging can be achieved with truly incoherent light [15], and then they worked under a particular path configuration, which is not the same as used in the quantum ghost image and interference experiments [1,2], and consequently they could not give the image formation equation and the interference fringe equation [15]. More recently, Cheng et al study the coincidence interference with a complete incoherent light by using classically theory [16], and as a matter of fact, the visibility of the fringes in their case is zero, that is to say, there are no interference fringes at all. In this paper, we study the ghost imaging by using a blackbody radiation source [17] and derive the image formation equation, and investigate the visibility and quality of the image and their dependence on the temperature and the surface size of the blackbody. The difference between the imaging with the entangled photon pairs and the imaging with blackbody radiation is also discussed.

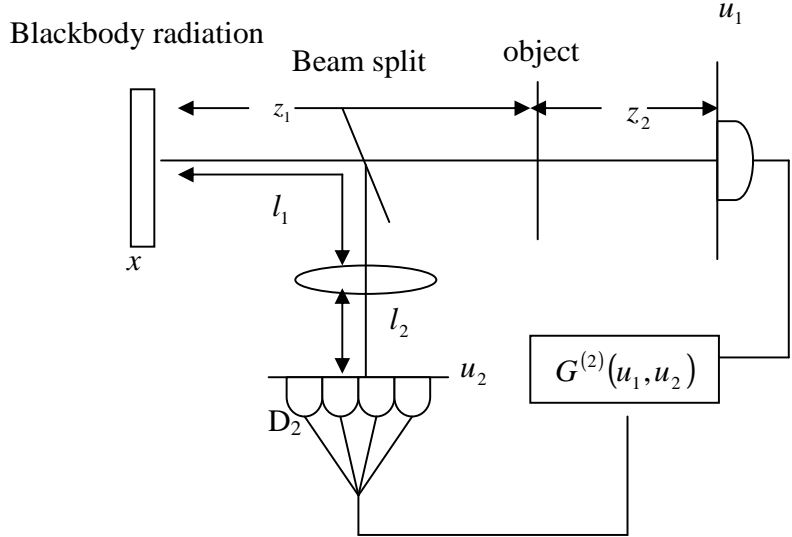

FIG.1 The scheme for ghost interference with blackbody radiation

The scheme for the ghost imaging with blackbody radiation is shown in Fig.1. The blackbody radiation first propagates through a beam split, then propagates through path one and two to detector one and two (D1 and D2). In path one (from the blackbody to D1), between the beam split and D1, there is an object (a double-slit aperture) with its transmision function $H(v)$. In path two (from the blackbody to D2), there is a lens with focal length $f$ between the beam split and detector two, and the distances from the blackbody to the lens and the lens to D2 are $l_1$ and $l_2$, respectively. The scheme is the same as in [1] except that the lens in parth two in order to emphasize the object and the image are in different paths. (The lens in path one will be discussed at the end.) The coincident counting rate is proportional to the fourth order correlation function, $G^{(2)}(u_1, u_2)$.

According the second order and fourth order optical coherence theory, the fourth order correlation function between the two detectors obeys the following integral formula for the blackbody radiation [17],

$$G^{(2)}(u_1,u_2) = \langle E(u_1)E(u_2)E^*(u_2)E^*(u_1)\rangle$$

$$= \int_{-\infty}^{\infty}\int_{-\infty}^{\infty}\int_{-\infty}^{\infty}\int_{-\infty}^{\infty} h_1(x_1,u_1)h_1^*(x_4,u_1)h_2(x_2,u_2)h_2^*(x_3,u_2)\times\langle E(x_1)E(x_2)E^*(x_3)E^*(x_4)\rangle dx_1dx_2dx_3dx_4$$

$$=<I(u_1)><I(u_2)>+|\Gamma(u_1,u_2)|^2, \qquad (1)$$

where $h_1(x_1,u_1)$, $h_2(x_2,u_2)$ are the response functions of the two paths through which the blackbody radiation passes, and

$$<I(u_i)>= \int_{-\infty}^{\infty}\int_{-\infty}^{\infty} h_i(x_1,u_i)h_i^*(x_2,u_i)\langle E(x_1)E^*(x_2)\rangle dx_1dx_2 \qquad i=1,2, \qquad (2)$$

$$\Gamma(u_1,u_2) = \int_{-\infty}^{\infty}\int_{-\infty}^{\infty}\langle E(x_1)E^*(x_2)\rangle h_1(x_1,u_1)h_2^*(x_2,u_2)dx_1dx_2. \qquad (3)$$

$<I(u_i)>$ is the second order correlation function at the same space point (the intensity at the i-th detector ), and depend only on the i-th path. $\Gamma(u_1,u_2)$ is the second order cross correlation function at two different points, which is related to both detectors and depends on both paths. Both $<I(u_1)><I(u_2)>$ and $\Gamma(u_1,u_2)$ have contribution to the coincident counting rates. In Eq. 1, the fourth order correlation is replaced by the second order correlation under the condition of $<E(x_i)>=0$ for the blackbody radiation [17].

For the blackbody radiation the second order correlation function takes the following form [17]:

$$\langle E_i(\mathbf{r}_1)E_j^*(\mathbf{r}_2)\rangle = \frac{1}{8\pi^3}\left(\frac{\hbar}{2\varepsilon_0}\right)\int\frac{\omega}{\exp(\hbar\omega/k_BT)-1}\left(\delta_{ij}-\frac{k_ik_j}{k^2}\right)\exp[i\mathbf{k}\cdot(\mathbf{r}_2-\mathbf{r}_1)]d^3k, \qquad (4)$$

where T is the blackbody's temperature, and $k_B$ is the Boltzmann's constant. For simplicity, we consider the source in one-dimension. The numerical calculation shows that $\langle E(x_1)E^*(x_2)\rangle$ is a quasi-Gaussian distribution over $(x_1-x_2)$ with a temperature dependent width [17]. Therefore, we can approximately write the second order correlation function as

$$\langle E(x_1)E^*(x_2)\rangle \approx I(x_1,\sigma_g)\exp\left(-\frac{x_1^2+x_2^2}{4\sigma_I^2}\right)\exp\left[\frac{(x_1-x_2)^2}{2\sigma_g^2}\right], \qquad (5)$$

where $\sigma_g$ is the correlation length and is temperature dependent. As the temperature goes to infinity, $\sigma_g$ approaches zero and $I(\mathbf{r}_1,\sigma_g)$ infinity. Here we introduced a Gaussian function $\exp\left(-\dfrac{x_1^2+x_2^2}{4\sigma_I^2}\right)$ to take into account the finite surface size of the blackbody with $\sigma_I$ the transverse size.

With the help of Collin's formula and the detail information of the two paths, $h_1(x_1,u_1)$ and $h_2(x_2,u_2)$ [18], we obtain

$$<I(u_1)>=\dfrac{1}{\lambda^2 z_1 z_2}\int_{-\infty}^{\infty}\int_{-\infty}^{\infty}\int_{-\infty}^{\infty}\int_{-\infty}^{\infty}<E(x_1)E(x_2)>H(v_1)H^*(v_2)\exp\left[-\dfrac{i\pi}{\lambda z_1}\left(x_1^2-2x_1v_1+v_1^2\right)\right]$$
$$\times\exp\left[-\dfrac{i\pi}{\lambda z_2}\left(v_1^2-2v_1u_1+u_1^2\right)\right]\exp\left[\dfrac{i\pi}{\lambda z_1}\left(x_2^2-2x_2v_2+v_2^2\right)\right]\exp\left[\dfrac{i\pi}{\lambda z_2}\left(v_2^2-2v_2u_1+u_1^2\right)\right]dx_1 dx_2 dv_1 dv_2. \quad (6)$$

For a fixed $u_1$, $<I(u_1)>$ is a constant, and it is easy to show that $<I(u_2)>=B_0=$const. The image pattern is determined by the cross correlation function,

$$\Gamma(u_1,u_2)=<E(u_1)E^*(u_2)>$$
$$=\left(-\dfrac{i}{\lambda^3 z_1 z_2 b_2}\right)^{1/2}\int_{-\infty}^{\infty}\int_{-\infty}^{\infty}\int_{-\infty}^{\infty}<E(x_1)E^*(x_2)>H(v_1)\exp\left[-\dfrac{i\pi}{\lambda z_1}\left(x_1^2-2x_1v_1+v_1^2\right)-\dfrac{i\pi}{\lambda z_2}\left(v_1^2-2v_1u_1+u_1^2\right)\right]$$
$$\exp\left[\dfrac{i\pi}{\lambda b_2}\left(a_2 x_2^2-2x_2 u_2+d_2 u_2^2\right)\right]\exp[-ik(z_1+z_2)+ik(l_1-l_2)]dx_1 dx_2 dv_1 \quad (7)$$

where $a_2, b_2, c_2$ and $d_2$ are the optical transfer matrix elements between the blackbody and detector two,

$$\begin{pmatrix}a_2 & b_2 \\ c_2 & d_2\end{pmatrix}=\begin{pmatrix}1 & l_2 \\ 0 & 1\end{pmatrix}\begin{pmatrix}1 & 0 \\ -1/f & 1\end{pmatrix}\begin{pmatrix}1 & l_1 \\ 0 & 1\end{pmatrix}=\begin{pmatrix}1-\dfrac{l_2}{f} & l_1+l_2-\dfrac{l_1 l_2}{f} \\ -\dfrac{1}{f} & 1-\dfrac{l_1}{f}\end{pmatrix}. \quad (8)$$

If we set

$$\dfrac{1}{l_1-z_1}+\dfrac{1}{l_2}=\dfrac{1}{f}, \quad (9)$$

Eq. (8) is reduced to

$$\begin{pmatrix}a_2 & b_2 \\ c_2 & d_2\end{pmatrix}=\begin{pmatrix}1 & l_2 \\ 0 & 1\end{pmatrix}\begin{pmatrix}1 & 0 \\ -\dfrac{1}{l_1-z_1}-\dfrac{1}{l_2} & 1\end{pmatrix}\begin{pmatrix}1 & l_1 \\ 0 & 1\end{pmatrix}=\begin{pmatrix}-\dfrac{l_2}{l_1-z_1} & -\dfrac{z_1 l_2}{l_1-z_1} \\ -\dfrac{1}{l_2}-\dfrac{1}{l_1-z_1} & 1-\dfrac{l_1}{l_2}-\dfrac{l_1}{l_1-z_1}\end{pmatrix} \quad (10)$$

By substituting Eqs. (5) and (10) into Eq. (7), we have the cross correlation function. In order to obtain an analytical expression, we assume blackbody is sufficiently large ($\sigma_I \to \infty$) and at very high temperature ($\sigma_g \to 0$), and we have

$$|\Gamma(u_1, u_2)|^2 = \frac{I_0^2}{\lambda z_2 |a_2|} \left| H\left(\frac{u_2}{a_2}\right) \right|^2. \quad (11)$$

This is a perfect image of the object with an amplification of $a_2$, which is obtained under Eq. (9), the Gaussian thin lens equation. We emphasize that the conditions ($\sigma_g \to 0$ and $\sigma_I \to \infty$) are only for obtaining the analytical expression of Eq. (11). When $\sigma_I$ and $\sigma_g$ are finite or Eq. (9) is not satisfied, the image needs to be obtained numerically. In Fig.2, we plot the evolution from the ghost image into an interference pattern of a double slits with slit width $a$, and distance of the two slits $d$, when we vary $l_2$ from satisfying to not satisfying Eq. (9). The transmission function of the double slits is $H(v) = 1$ for $-\frac{d}{2} - \frac{a}{2} < v < -\frac{d}{2} + \frac{a}{2}$ and $\frac{d}{2} - \frac{a}{2} < v < \frac{d}{2} + \frac{a}{2}$ and $=0$ otherwise,

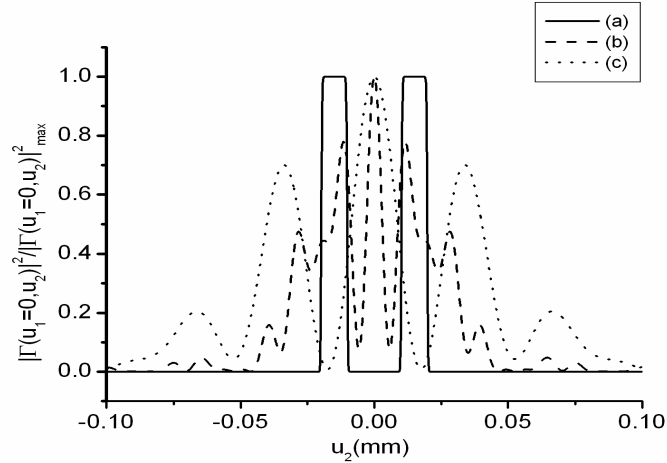

FIG.2 The image pattern of a double slit in the scheme of Fig.2 with blackbody radiation for different $l_2$. (a) $l_2 = 20mm$ (imaging case), (b) $l_2 = 20.5mm$, (c) $l_2 = 21.5mm$ with $\lambda = 702nm$, $a = 0.01mm$, $d = 0.03mm$, $z_1 = 10mm$, $z_2 = 40mm$, $f = 10mm$, $l_1 = 30mm$ and $\sigma_I = 5mm$, $\sigma_g = 0.00001$.

Now we need to consider the visibility of the image. We define the visibility of the image as

$$V = \frac{Max|\Gamma(u_1=0,u_2)|^2}{Max\langle I(u_1=0)\rangle\langle I\langle u_2\rangle\rangle},\qquad(12)$$

Under the condition of obtaining Eq. (11), $\sigma_g \to 0$ and $\sigma_I \to \infty$, the visibility of the image is zero. The images of the double slit aperture for different transverse sizes and transverse coherence widths (temperature) of the blackbody are shown in Fig.3 and Fig. 4, respectively. From Fig.3, we can find that when the source's transverse size $\sigma_I$ increases, the quality of the image increases, while the visibility decreases. From Fig.4, we can find that when the source's transverse coherence width $\sigma_g$ decreases, quality of the image increases, while the visibility decreases.

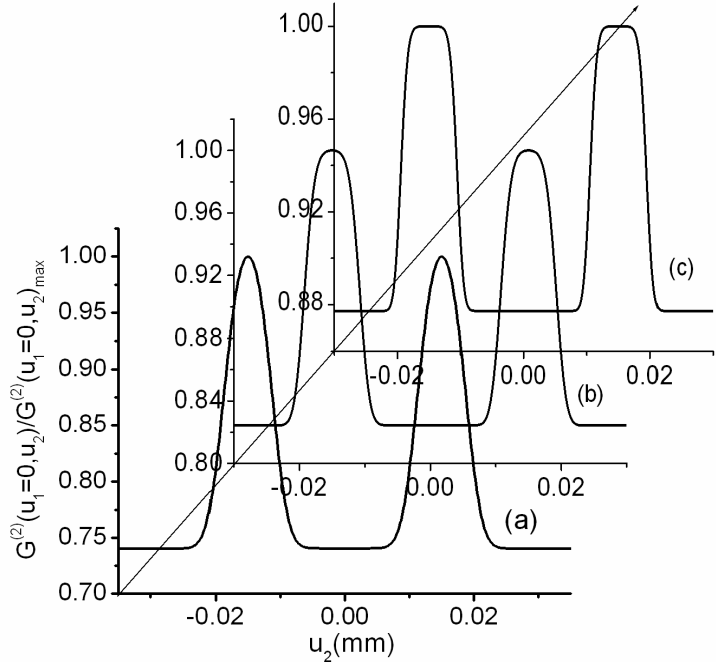

Fig.3 The image of the double slit aperture for different source's transverse size $\sigma_I$: (a) 0.1mm, (b) 1mm and (c) 5mm with $\lambda = 702nm$, $a = 0.01mm$, $d = 0.03mm$, $z_1 = 10mm$, $z_2 = 40mm$, $f = 10mm$, $l_1 = 30mm$ and $l_2 = 20mm$, $\sigma_g = 0.001$mm.

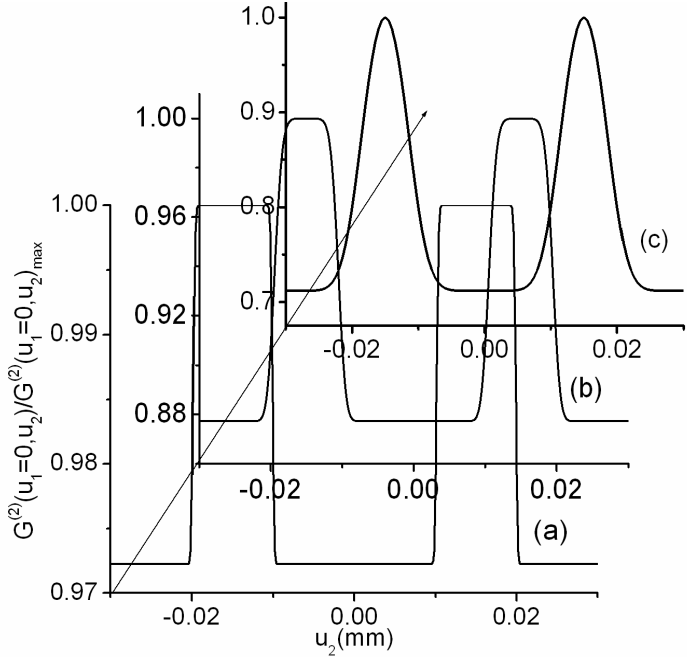

Fig.4 The image of a double slit aperture for different source's transverse coherence width $\sigma_g$: (a) 0.00001mm, (b) 0.0005mm and (c) 0.003mm with $\lambda = 702nm$, $a = 0.01mm$, $d = 0.03mm$, $z_1 = 10mm$, $z_2 = 40mm$, $f = 10mm$, $l_1 = 30mm$ and $l_2 = 20mm$, $\sigma_I = 5mm$.

Now we study the dependence of the visibility and quality of the image on the source's transverse coherence width $\sigma_g$ and transverse size $\sigma_I$ in details. We define an image quality factor,

$$Q = \frac{\int_{-\infty}^{\infty} \left| \frac{|\Gamma(u_1 = 0, u_2)|^2}{|\Gamma(u_1 = 0, u_2)|^2_{max}} - |H(u_2)|^2 \right| du_2}{\int_{-\infty}^{\infty} |H(u_2)|^2 \, du_2}. \qquad (13)$$

Small $Q$ value corresponds to high image quality. The dependences of the visibility and quality of the image of the double slit aperture on the transverse coherence width are shown in Fig.5 and Fig.6, respectively. High quality is companied by poor visibility, and good visibility companied by low quality. In order to observe the classical ghost image with blackbody radiation, the selection of suitable transverse size and transverse coherence width is essential.

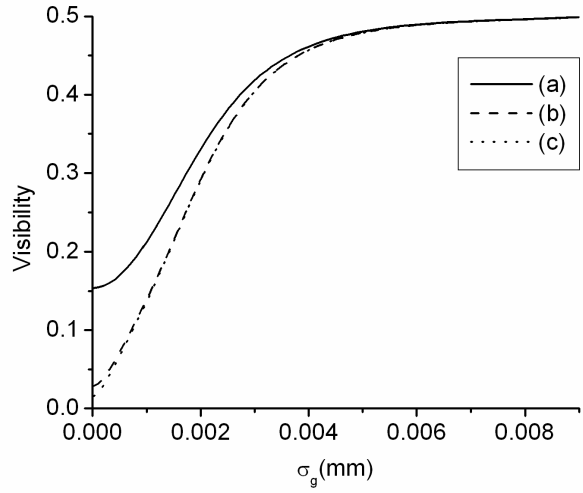

Fig.5. Evolution of the visibility of the image of a double slit aperture versus transverse coherence width $\sigma_g$ for different transverse sizes of the source, $\sigma_I$. (a) 1mm (b) 5mm (c) 10mm with $\lambda = 702nm$, $a = 0.01mm$, $d = 0.03mm$, $z_1 = 10mm$, $z_2 = 40mm$, $f = 10mm$, $l_1 = 30mm$ and $l_2 = 20mm$.

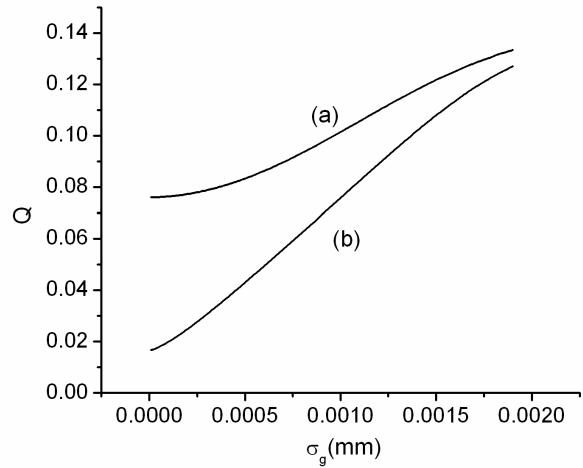

Fig.6. Evolution of the quality of the image of a double slit aperture versus transverse coherence width $\sigma_g$ for different transverse sizes of the source, $\sigma_I$. (a) 1mm (b) 5mm with $\lambda = 702nm$, $a = 0.01mm$, $d = 0.03mm$, $z_1 = 10mm$, $z_2 = 40mm$, $f = 10mm$, $l_1 = 30mm$ and $l_2 = 20mm$.

Now, we ask ourselves whether the classical ghost image is the same as the quantum ghost image with the entangled photon pairs. Comparing Eqs. (9) with the corresponding equations for the quantum ghost image [2], the following differences have been found: (1) $l_1 + z_1$ in the imaging formation equation for the quantum case is replaced by $l_1 - z_1$ for the blackbody. (2) For $l_1 > z_1$, we have an inverted image and enlarged image with $l_2 > 2f$ or an inverted and reduced image with $2f > l_2 > f$; for $l_1 < z_1$, we have an erect and reduced image with $0 < l_2 < f$. For the quantum ghost image, the image is always inverted. (3) Another important difference between quantum and classical ghost image is the visibility. In the Quantum ghost image, good quality and high visibility can be achieved simultaneously [1, 2], but this is impossible in the ghost image with blackbody radiation. The differences come from the classical and quantum nature of the correlation functions. In quantum case, the cross correlation function is $\Gamma(x_1, x_2) \propto \langle 0,0 | E(x_1) E(x_2) | \Psi \rangle$ due to the entanglement, while in the classical case (no entanglement) it is Eq. (3) where we have $E^*(x_2)$ instead of $E(x_2)$. If a 50% phase conjugate mirror is used for the beam split, we will have $(l_1 - z_0) + (z_1 - z_0)$ instead of $(l_1 - z_0) - (z_1 - z_0)$ with $z_0$ the distance between the black body and the beam split. If the lens is in path one, we will have $\frac{1}{S_1 - S_2} + \frac{1}{S_3} = \frac{1}{f}$ with $S_1$ and $S_2$ the distances from the blackbody to the lens and to D2, and $S_3$ from the lens to the object. (In quantum case, $-S_2$ is replaced by $+S_2$.) Further more, in quantum case we have $G^{(2)}(x_1, x_2) = |\Gamma(x_1, x_2)|^2$, that is to say, there is no background noise and consequently the visibility is high. In the classical case the background noise limits the visibility.

If we take out the lens in path two (see Fig. 1), we have the ghost interference. With the same calculation above, we can obtain the same interference fringes as obtained from the quantum ghost interference except the replacement of $-z_1$ by $+z_1$.

In conclusion, we have invested the ghost image created with blackbody radiation by using the optical coherence theory. The ghost image formation depends on both paths. To obtain the ghost image, a Gaussian thin lens equation must be satisfied. The ghost image is gradually blurred out, when the temperature (or the size) decreases. The quality of the ghost image increases with the increase of the temperature and the size of the blackbody. As the temperature and the size both increase to infinity, we will have a perfect image, but the visibility goes to zero. The nature of the ghost imaging is due to the entanglement in the entangled photon pair and is due to the Hanbury Brown-Twiss Effect (low coherence with fluctuations but not completely no coherence) in the blackbody radiation.

**Acknowledgement:** This work was supported by RGC (CA02/03.SCI).